\documentclass{article}
\usepackage{spconf,amsmath,graphicx,amssymb,url}
\usepackage{caption}
\usepackage{subcaption}
\usepackage{microtype}
\usepackage{hyperref}
\newcommand{\Lc}{L_\text{c}}

\usepackage{color}
\definecolor{purple(x11)}{rgb}{0.63, 0.36, 0.94}
\definecolor{cadmiumgreen}{rgb}{0.0, 0.42, 0.24}

\title{Joint UE positioning and distributed sensing in the upper mid-band exploiting virtual  apertures}
\name{Soroush Mesforush, Murat Bayraktar, and Nuria Gonz\'alez-Prelcic}
\address{University of California San Diego, La Jolla, CA, USA}
\usepackage{caption}
\begin{document}
% \ninept
%
\maketitle
\begin{abstract}
Networks exploiting distributed integrated sensing and communication (DISAC) nodes can provide enhanced localization and sensing performance, further emphasized when operating with  large arrays and bandwidths available in the upper mid-band (also known as FR3). In this paper, we consider a DISAC system operating at FR3 where  a single base station (BS) acts as the transmitter and several vehicular user equipments (UEs) act as the receivers. We tackle the design of the signal processing chain at the UE side to enable joint UE positioning and target localization. The system model exploits a multiple-input-multiple-output orthogonal frequency division multiplexing (MIMO-OFDM) waveform, and incorporates practical effects such as inter-node timing offsets (TOs), extended targets, dense multipath, and realistic uniform planar arrays (UPAs) at both ends. 
The proposed design includes a  multipath estimation stage at each UE, clutter removal,  a novel clustering and association scheme, and a final joint estimator of UE positions and target locations. The estimator solves a weighted least squares (WLS) problem to jointly compute clock offsets and localize UEs and targets. Numerical results considering two UEs and two targets show that for 80\% of the cases the target localization error is below 32cm, while the UE positioning error is below 44cm.  
\end{abstract}

\vspace{-4mm}
\section{Introduction}
\vspace{-2mm}
%Motivate the interest of DISAC versus ISAC and identify challenges in a general way
Unlike single-node ISAC, DISAC networks which exploit multi-node coordination can jointly deliver connectivity and multi-perspective sensing \cite{Strinati2025DISAC, Guo2025DMIMO}. When distributed operation is combined with the exploitation of large arrays and bandwidths, as those available in the upper midband, it is theoretically possible to achieve a very high accuracy for UE positioning and target localization \cite{Meng2024CoopISAC}. 
Many challenges arise, however, when designing the signal processing chain in DISAC systems, mainly created by tight synchronization requirements for coherent processing, hardware impairments that require calibration, and the complexity of the extended scenario in many practical settings and use cases.

%Coherent vs. incoherent DISAC depending on level of synchronization available. Prior work on incoherent DISAC. How much better we can do with coherent DISAC and those virtal apertures. Limited prior work   on coherent DISAC. 
Depending on the level of synchronization available, DISAC systems will exploit incoherent or coherent  approaches for the design of the sensing processing chain.  Prior work on DISAC has focused on incoherent processing strategies, which do not require tight synchronization in time, frequency or phase \cite{han2025beamforming,han2025signalingdesignnoncoherentdistributed}. Cooperative formulations using factor graphs and/or Bayesian fusion improve the geometry and NLoS robustness, however, these methods often increase computational burden or retain the tight timing assumptions and typically emphasize long timing horizons \cite{wymeersch2009cooperative,10584443}. %In particular, \cite{wymeersch2009cooperative} notes that time of arrival (ToA) formulations widely presume tight synchronization whereas round trip time of arrival (ToA) avoids a common clock between nodes. 
Coherent DISAC augments single-node ISAC by fusing measurements from spatially separated nodes into a virtual aperture, boosting geometric diversity and the observability of time/angle cues \cite{Zhang2024CoopISAC}. To tackle asynchronism challenges for bistatic/multistatic settings and enable coherence, a few previous designs have explicitly accounted for timing offset (TO) estimation, mostly under simplified channels or point scatterer modeling \cite{pegoraro2024jump,10769985}. Over the air synchronization can also make phase aligned fusion across nodes possible, as shown in \cite{han2025over}. 

% Contributions of this paper
In this paper, we propose the first single shot DISAC system for joint UE positioning and target localization operating at FR3. A base station (BS) equipped with UPAs serves as the sole transmitter in the system, while several UEs--also equipped with UPAs--operate as the distributed receivers. Unlike previous work, we consider  a realistic channel explicitly including extended target scattering,  realistic multipath communication channels between the BS and the UEs generated by ray tracing \cite{remcom2025wireless}, unknown UE positions and inter-node TOs.  After preprocessing at each UE, %At each UE, we exploit high resolution beamspace ESPRIT  to estimate  per path parameters such as angles, delays, and gains from tensorized measurements. Each UE performs angle domain prefiltering using a fixed azimuth–elevation mask to remove clutter. We also design a novel clustering and matching method for unsupervised grouping of echoes, simultaneously aligning clusters across asynchronous vehicles without prior target count is designed. 
the fusion center (FC) stacks spatial and delay measurements from all UEs to formulate a weighted least squares 
(WLS) problem to obtain UE positions, target locations, and TOs. 
Numerical evaluations shows that DISAC with collaborating UEs outperforms single-UE ISAC systems, while simulations over realistic upper mid-band channels demonstrate dm-level accuracy even with a single snapshot and imperfect synchronization

\vspace*{-4mm}
\section{System Model}
\begin{figure}[t!]
	\begin{center}\includegraphics[width=\linewidth]{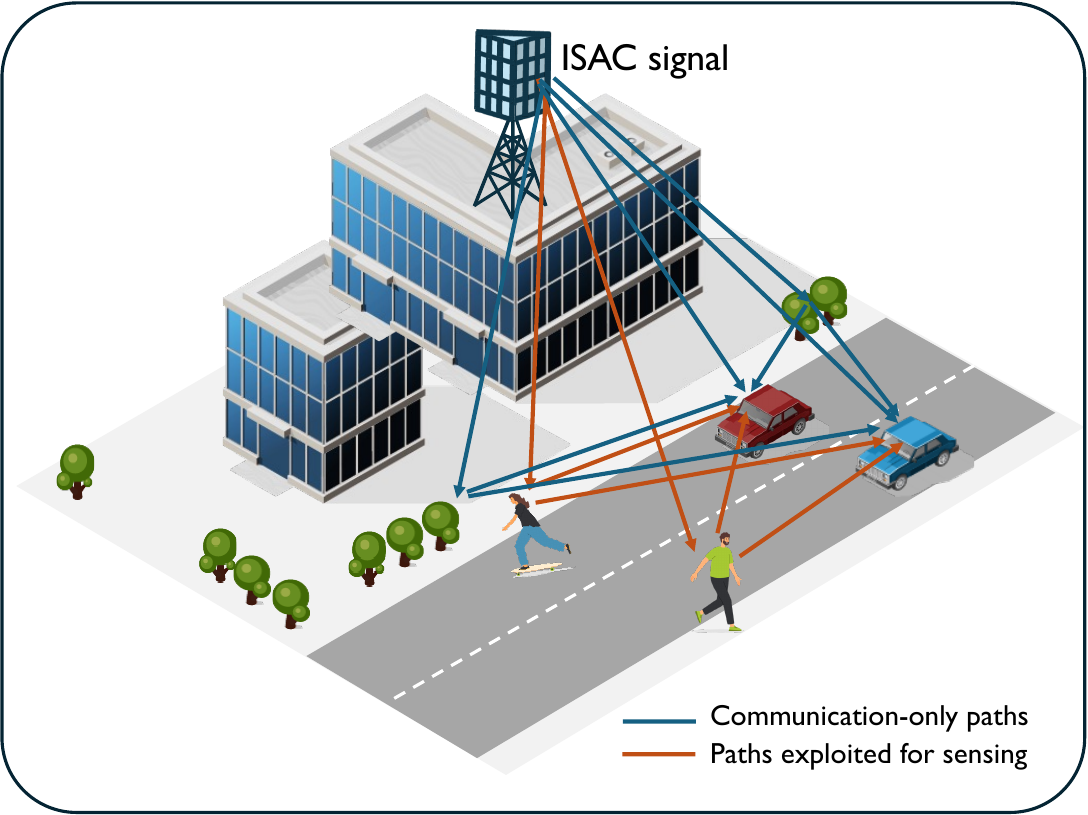}
		\caption{DISAC system where a BS acts as transmitter and several vehicular UEs of unknown positions act as the receivers in a multistatic sensing setting.  }
\label{fig:sysmod}	
\end{center}
\end{figure}

\vspace*{-2mm}
We consider the downlink of a cellular  system supporting $N$ vehicular users (UEs) in an urban environment and operating in the upper midband. The BS is
mounted on the top of a building and operates with a large uniform planar array (UPA), as illustrated in  \ref{fig:sysmod}. The UEs are also equipped with UPAs of a smaller size. The number of non-connected targets in the environment is $Q$.  The system is based on a MIMO-OFDM signal, with  $I$ OFDM pilot symbols and $K$ subcarriers exploited for downlink channel estimation and single shot initial localization of the targets and the UEs. The frequency domain channel model between the BS and a UE, at subcarrier $k$ and OFDM symbol $i$--accounting for extended targets, TO, and dense multipath--is written as
\vspace*{-2mm}
\begin{equation}\label{eq:channel_model}
\begin{split}
\mathbf{H}_{k,i} &= \sum_{c=0}^{C-1}\sum_{\ell=0}^{\Lc-1} b_{c,\ell} e^{-j2\pi\,k(\tau_{c,\ell} -t_i)\Delta f} \\
&\quad \times \mathbf{a}_{\mathrm{R}}(\theta_{c,\ell},\phi_{c,\ell})\, \mathbf{a}_{\mathrm{T}}^\mathrm{H}(\vartheta_{c,\ell},\varphi_{c,\ell}) 
%&= \sum_{c=0}^{C-1}e^{j2\pi\,k\Delta f\,t_i} \sum_{\ell=0}^{L_c-1} b_{c,\ell} e^{-j2\pi\,k\Delta f\,\tau_{c,\ell}} \\
%&\quad \times \mathbf{a}_{\mathrm{R}}(\theta_{c,\ell},\phi_{c,\ell})\, \mathbf{a}_{\mathrm{T}}^\mathrm{H}(\vartheta_{c,\ell},\varphi_{c,\ell}),
\end{split}
\end{equation}
where $C$ is the number of clusters, $\Lc$ is the number of delay taps, the  $\ell$-th path  in the $c$-th  cluster has complex gain $b_{c,\ell}$, delay $\tau_{c,\ell}$, angles of arrival (AoAs) $(\theta_{c,\ell},\phi_{c,\ell})$, and angles of departure (AoDs) $(\vartheta_{c,\ell},\varphi_{c,\ell})$, with $\theta,\vartheta$ denoting elevation and $\phi,\varphi$ denoting azimuth. The steering vectors at the BS and the UE are denoted as $\mathbf{a}_{\mathrm{T}}$ and $\mathbf{a}_{\mathrm{R}}$, respectively.  The TO between the BS and the $i$-th UEs  is denoted as $t_i$ and assumed to be similar across all paths and constant for all OFDM symbols during the coherence time. 
For first order paths, the delay can be computed as
\begin{equation}\label{eq:delay_eq}
\tau_{c,\ell}= \frac{\|\mathbf{p}_{c,\ell} - \mathbf{p}_{\text{BS}}\| + \|\mathbf{p}_{\mathrm{v}} - \mathbf{p}_{c,\ell}\|}{v_c}-t_i ,
\end{equation}
where $\mathbf{p}_{\mathrm{v}}$ and $\mathbf{p}_{\mathrm{BS}}$ are the UE and BS positions, $\mathbf{p}_{c,\ell}$ is the location of the scattering point for that path,  and $v_c$ is the speed of light. The steering vector for a UPA is defined as $\mathbf{a}_\mathrm{UPA}(\gamma,\eta) = \mathbf{a}_\mathrm{x}(\gamma,\eta)\otimes \mathbf{a}_\mathrm{y}(\gamma,\eta)$; where $\mathbf{a}_x(\gamma,\eta) =\big[e^{\,j k d\, n\,\sin\gamma\cos\eta}\big]_{n=0}^{N_x-1}$, and $\mathbf{a}_y(\gamma,\eta) = \big[e^{\,j k d\, m\,\sin\gamma\sin\eta}\big]_{m=0}^{N_y-1}$. where $N_x$ and $N_y$ are the number of antennas along the $x-$ and $y-$ axes, $k=\frac{2\pi}{\lambda}$ is the wavenumber, $d$ is the spacing between elements, and $\gamma$, and $\eta$ are azimuth and elevation angles. 

The receive signal at the $n$-th UE, $k$-th subcarrier and $i$-th  OFDM symbol is 
\begin{equation}\label{eq:received_signal}
\mathbf{y}_{k,i}^{(n)}
= \bigl[{\mathbf W}_{k,i}^{(n)}\bigr]^{\mathrm{H}}
  \mathbf{H}_{k,i}^{(n)}
  \mathbf{F}_{k,i}\,\mathbf{d}_{k,i} 
  + \bigl[{\mathbf W}_{k,i}^{(n)}\bigr]^{\mathrm{H}}\,\mathbf{z}_{k,i}^{(n)},
\end{equation}
where  $\mathbf{W}_{k,i}^{(n)} \in \mathbb{C}^{N_\mathrm{R} \times N_\mathrm{s}}$ is the hybrid combining matrix used for training, where $N_\mathrm{R}$ is the number of receive antennas and $N_\mathrm{s}$ is the number of processed streams. The channel $\mathbf{H}_{k,i}^{(n)} \in \mathbb{C}^{N_\mathrm{R} \times N_\mathrm{T}}$, is modeled as in \eqref{eq:channel_model}, with $N_\mathrm{T}$ transmit antennas at the BS. At the transmitter, the pilot signal $\mathbf{d}_{i,k} \in \mathbb{C}^{N_{\text{s}}}$ is shaped by a training  hybrid precoding matrix $\mathbf{F}_{k,i} \in \mathbb{C}^{N_\mathrm{T} \times N_{\text{s}}}$. The additive noise, $\mathbf{z}_{k,i}^{(n)}$, is assumed to be a complex Gaussian random variable, $\mathcal{N_C}(\mathbf{0},\sigma^2_z\mathbf{I}_{N_\mathrm{R}})$. The training precoders and combiners are created using a DFT codebook. The specific design of these hybrid beamforming matrices follows the methodology presented in \cite{10623018}. At each UE, channel estimation is performed using the received signal using the aforementioned UE pilots. Estimated parameters are passed on to a UE serving as the FC, where joint UE positioning, target localization and timing offset estimation is implemented.

% This received signal is exploited at each UE for independent channel estimation and  at the fusion center for joint target and UE localization. 

%\vspace{-.3cm}
\section{Joint UE and Target Localization and TO Estimation}
\begin{figure}[h]
\begin{center}\includegraphics[width=\linewidth]{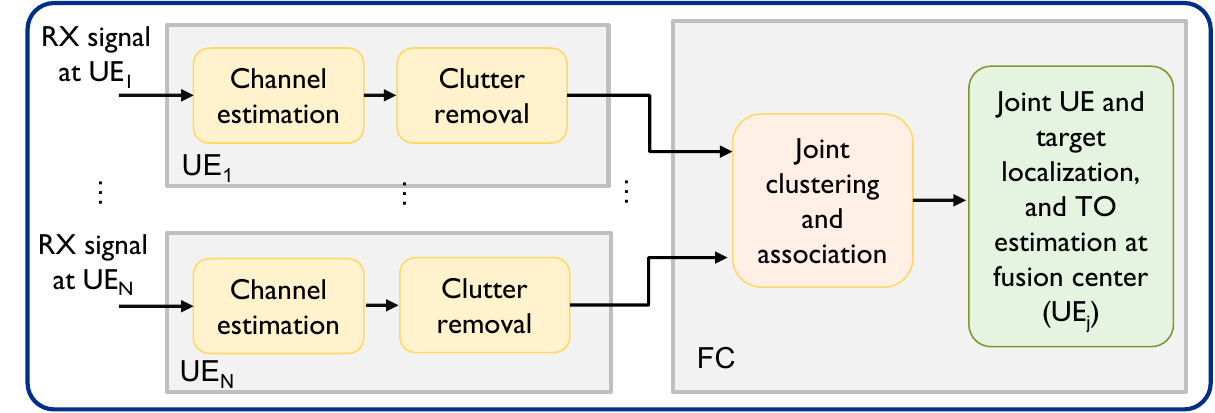}
\caption{Distributed processing chain for joint UE and target localization and TO estimation at a fusion center.}\label{fig:2}	
\end{center}
\end{figure}

In this section, we describe the proposed distributed pipeline. First, we formulate and solve the multipath estimation problem. Second, we explain the logic behind the clutter removal stage. Third, we design a novel joint clustering and association step. Finally we describe the joint UE and target localization and TO estimation step at the FC. An illustration of the distributed chain is shown in Fig. \ref{fig:2}.

\vspace{-5mm}
\subsection{Channel Estimation via  Beamspace ESPRIT}
Beamspace ESPRIT builds on the canonical polyadic (CP) decomposition of an order-$O$ tensor $\mathbf{\Upsilon}\in\mathbb{C}^{N_1\times N_2\times\cdots\times N_O}$, that can be written as
\begin{equation}\label{eq:CPInit}
    \mathbf{\Upsilon} = \sum_{\ell=1}^{N_L}\gamma_\ell \mathbf{u}_{1,\ell}\circ \mathbf{u}_{2,\ell}\circ\cdots\circ \mathbf{u}_{O,\ell},
\end{equation}
where $\mathbf{u}_{o,\ell}$ is the rank-1 tensor spanning dimension $o$ of the $\ell$-th tensor product and $\gamma_\ell$ is the coefficient of the appropriate tensor product. The kernel of beamspace ESPRIT is based on solving the following optimization problem \cite{10623018}
\begin{equation}\label{eq:OPTCPD}
\min_{\hat{\mathbf{U}}_o}  \
\Big\|\mathbf{\Upsilon}-\sum_{\ell=1}^{N_L}
\hat{\mathbf u}_{1,\ell}\,\circ\cdots\circ\,\hat{\mathbf u}_{O,\ell}
\Big\|_\mathrm{F}.
\end{equation}
To apply beamspace ESPRIT to the channel estimation problem, we rewrite \eqref{eq:received_signal} for one UE as
\begin{equation}
\label{eq:beamspace-snapshot}
\mathbf{y}_{k,i} \;\triangleq\; \mathbf{W}_{k,i}^{\mathrm{H}}\mathbf{H}_{k,i}\mathbf{F}_{k,i}\,\mathbf{d}_{k,i} \;+\; \mathbf{W}_{k,i}^{\mathrm{H}}\mathbf{z}_{k,i},
\end{equation} 
where the matrices $\mathbf{W}_{k,i} \;=\; \mathbf{W}_{\mathrm{az}}\otimes \mathbf{W}_{\mathrm{el}}$ and $\mathbf{F}_{k,i} \;=\; \mathbf{F}_{\mathrm{az}}\otimes \mathbf{F}_{\mathrm{el}}$ represent the hybrid combining and precoding, respectively, where $\otimes$ is the Kronecker product. The receive combiners are $\mathbf{W}_{\mathrm{az}}\in\mathbb{C}^{N_{v,x}\times M_{\mathrm{R,az}}}$ and $\mathbf{W}_{\mathrm{el}}\in\mathbb{C}^{N_{v,y}\times M_{\mathrm{R,el}}}$. Similarly, the transmit beamformers are $\mathbf{F}_{\mathrm{az}}\in\mathbb{C}^{N_{\mathrm{B},x}\times M_{\mathrm{T,az}}}$ and $\mathbf{F}_{\mathrm{el}}\in\mathbb{C}^{N_{\mathrm{B},y}\times M_{\mathrm{T,el}}}$. The dimensions are defined by the number of UPA elements along the $x$- and $y$-axes at the receiver ($N_{v,x}, N_{v,y}$) and transmitter ($N_{B,x}, N_{B,y}$), and the number of beams along the azimuth and elevation dimensions for reception ($M_{\mathrm{R,az}}, M_{\mathrm{R,el}}$) and transmission ($M_{\mathrm{T,az}}, M_{\mathrm{T,el}}$). We form the 5-dimensional tensor with each dimension corresponding to AoAs (in azimuth and elevation), AoDs (in azimuth and elevation), and delays as
\vspace{-2mm}
\begin{equation}\label{eq:5D}
\begin{aligned}
\mathbf{Y}
&= \sum_{\ell=1}^{L} \gamma_\ell\,
   \mathbf{W}_{\mathrm{el}}^{*}\mathbf{a}(\theta_\ell)
   \circ \mathbf{W}_{\mathrm{az}}^{*}\mathbf{a}(\phi_\ell)
   \circ \mathbf{F}_{\mathrm{el}}^{*}\mathbf{a}(\vartheta_\ell)\\
&\quad \circ \mathbf{F}_{\mathrm{az}}^{*}\mathbf{a}(\varphi_\ell)
   \circ \mathbf{s}(\tau_\ell)
   + \mathbf{N},
\end{aligned}
\end{equation}
 where  $\gamma_\ell$ is the complex gain for path $\ell$. The OFDM Vandermonde vector is defined as $\mathbf{s}(\tau_\ell) = \left[e^{-j2\pi\Delta f\tau_\ell k}\right]_{k=0}^{K-1}$. The combined noise on the $k$-th subcarrier and $i$-th OFDM symbol, $\mathbf{n}_{k,i}=\mathbf{W}_{k,i}^\mathrm{H} \mathbf{z}_{k,i}$, follows a complex Gaussian distribution $\mathcal{N_C}(\mathbf{0},\sigma_z^2 \mathbf{W}^\mathrm{H}_{k,i}\mathbf{W}_{k,i})$. Consequently, the beamspace noise tensor $\mathbf{N}$ follows $\mathrm{cov}\!\big\{\mathrm{vec}(\mathbf N)\big\}
=\sigma_z^2\,
\big(\mathbf I_{K}\otimes \mathbf I_{M_{\mathrm{T,az}}}\otimes \mathbf I_{M_{\mathrm{T,el}}}
\otimes \mathbf{W}_{\mathrm{az}}^{\!\mathrm{H}}\mathbf W_{\mathrm{az}}\otimes\mathbf W_{\mathrm{el}}^{\!\mathrm{H}}\mathbf W_{\mathrm{el}}\big)$. To estimate the paths, we solve problem  \eqref{eq:OPTCPD} substituting the tensor with  $\mathbf{Y}$ in \eqref{eq:5D}.

\vspace{-3mm}
\subsection{Clutter Removal}
\vspace{-1mm}
ESPRIT-based channel estimation provides the delays, gains, AoAs, and AoDs for all paths. Given that the target is modeled as an extended object producing multiple reflections, it is essential to associate the angular components with their respective physical targets. 
However, environmental clutter can significantly degrade performance if not removed. To mitigate this, we exploit the fact that each vehicle primarily detects targets in its forward-looking direction. Accordingly, we apply a filtering process that retains only the paths with an angle of arrival (AoA) within a designated field of interest (FoI): 
\begin{equation}
-\phi_{\rm r}\leq\phi\leq\phi_{\rm r},\quad-\theta_{\rm r}\leq\theta\leq \theta_{\rm r}
\end{equation}
where $\theta_{\rm r}$ and $\phi_{\rm r}$ denote the angular bounds.

\vspace{-4mm}
\subsection{Joint Clustering and Association}\label{sub:Clus}
\vspace{-2mm}
After eliminating clutter, it is crucial to group and label all  paths  originating from the same target (clustering). Moreover, to have target-level consistency, we must determine the correspondence between a specific target identified at one UE and the same target observed at other UEs (association).  %To do so, per UE, we aim to solve the joint UE positioning, target localization, and TO estimation per path. 
We propose next a novel clustering and association procedure adapted to the DISAC scenario with sensing UEs. 
We treat each measured path at a given UE as if it was originated from a separate point target. Then, we form the following spatial equations:
\vspace{-3mm}
\begin{equation}\label{eq:spatialLocal}
\begin{aligned}
&\mathbf{u}_{\mathrm{BS}}^{(m,\ell)}\,r_m
+\mathbf{u}_{\mathrm{v}}^{(m,\ell)}\,d_{m}
  -\mathbf{p}_{\mathrm{v}}=-\mathbf{p}_\text{BS}\\
  & \mathbf{p}_{\mathrm{v}} = \mathbf{p}_{\text{BS}} + r^\text{LoS}\mathbf{u}^\text{LoS},
\end{aligned}
\end{equation}
where $\mathbf{u}_{\mathrm{BS}}^{(m,\ell)}$ and $\mathbf{u}^\text{LoS}$ are the unit direction vectors between the BS and the $l$-th scattering point of the $m$-th target and the LoS to the UEs,  respectively. Furthermore, $r_m$ is the BS-$m$-th target range, $d_{m}$ is the UE-$m$-th target range,  and $r^{\mathrm{LoS}}$ is the BS-UE LoS range; the unknowns are $\mathbf{p}_{\mathrm v}$ and these ranges. Similarly, we can write the following equations exploiting the delays:
\begin{equation}\label{eq:delaylocal}
    \begin{aligned}
&r_m + d_{m} + v_c\,\Delta t = v_c\,\tau_{m,\ell}\\
 &r^\text{LoS} +v_c\Delta t = v_c\tau^{\text{LoS}},
    \end{aligned}
\end{equation}
where $\Delta t$ is the  UE TO, and $\tau_{m,\ell}$ and $\tau^{\mathrm{LoS}}$ denote the NLoS and LoS delays, respectively. Note that in \eqref{eq:spatialLocal} and \eqref{eq:delaylocal}, since the equations are for a given UE, the UE index $i$ is omitted for brevity. The WLS equation and its solution can be written as
\begin{equation}\label{eq:WLS}
\min_{\mathbf{x}} \|\mathbf{W}^\frac{1}{2}(\mathbf{A}\mathbf{x} - \mathbf{b})\|^2 \longrightarrow \hat{\mathbf{x}} = \left( \mathbf{A}^\mathrm{T} \mathbf{W} \mathbf{A} \right)^{-1} \mathbf{A}^\mathrm{T} \mathbf{W} \mathbf{b},
\end{equation}
where $\mathbf{A}$ is built by stacking the linear coefficients from all spatial \eqref{eq:spatialLocal} and delay equations \eqref{eq:delaylocal}, and $\mathbf{b}$ is constructed by stacking the known components of the aforementioned equations. The WLS equation is solved to perform per-UE localization of the separate point targets. After this, the localized points per UE are fed to the DBSCAN algorithm \cite{ester1996dbscan}, which groups paths associated to the same target. This algorithm is appropriately tuned to identify individual reflections
as clusters if necessary, while also omitting spurious detections such as shadow targets. By doing this, clustering and matching is performed simultaneously and all measurements are in correspondence with their respective targets in the correct ordering required at the final localization stage.
\vspace{-3mm}
\subsection{Joint UE and Target Localization and TO Estimation at the Fusion Center}
After clustering and associating the per-UE measurements, they are sent to the fusion center (FC), where joint UE and target localization together with TO estimation is performed using the measurements from all UEs. The general steps and geometric equations are similar to those in Sec.~\ref{sub:Clus}, but are modified to make use of the information from all UEs to localize the same point for extended targets. Furthermore, a scenario with $N$ UEs and $M$ targets is considered. This way,  the vector of unknowns is defined as
\vspace{-2mm}
\begin{equation}\label{eq:Paramsx}
\begin{aligned}
\mathbf{x} = \bigl[
  &\,r_{1},\ldots,r_{M},\;
    d_{1,1},\ldots,d_{M,N},\;
    \Delta t_{1},\ldots,\Delta t_{N}, \\
  &\,x_{v,1},y_{v,1},z_{v,1},\ldots,
    x_{v,N},y_{v,N},z_{v,N},\\
    &
    r_{\mathrm{LoS},1},\ldots,r_{\mathrm{LoS},N}
\bigr]^{\mathrm{T}}
\end{aligned}
\end{equation}
where $\mathbf{x}\in\mathbb{R}^{M+MN+N+3N+N}$. The spatial and delay equations follow the general formulation in  \eqref{eq:spatialLocal} and \eqref{eq:delaylocal}. However, since extended targets and multiple UEs are considered, $\mathbf{u}_{\mathrm{v}}^{(m,i,\ell)}$, and $\tau_{m,i,\ell}$ must be used instead of $\mathbf{u}_{\mathrm{v}}^{(m,\ell)}$, and $\tau_{m,\ell}$ to account for the dependence with the UE index. By stacking all the equations together and defining the weight matrix as the diagonal matrix with the absolute values of the complex gains, i.e.  $\mathbf{W}=\text{diag}(|\gamma_1|,...,|\gamma_{4MNL+4N}|)\in\mathbb{R}^{4MNL+4N\times4MNL+4N}$, we create a WLS problem as in \eqref{eq:WLS} that can be solved to perform joint UE and target positioning and TO estimation.

\vspace*{-5mm}
\section{Simulation results}
\begin{figure}[t!]
  \centering
  \subfloat[\label{fig:ray}]{%
    \includegraphics[width=0.24\textwidth]{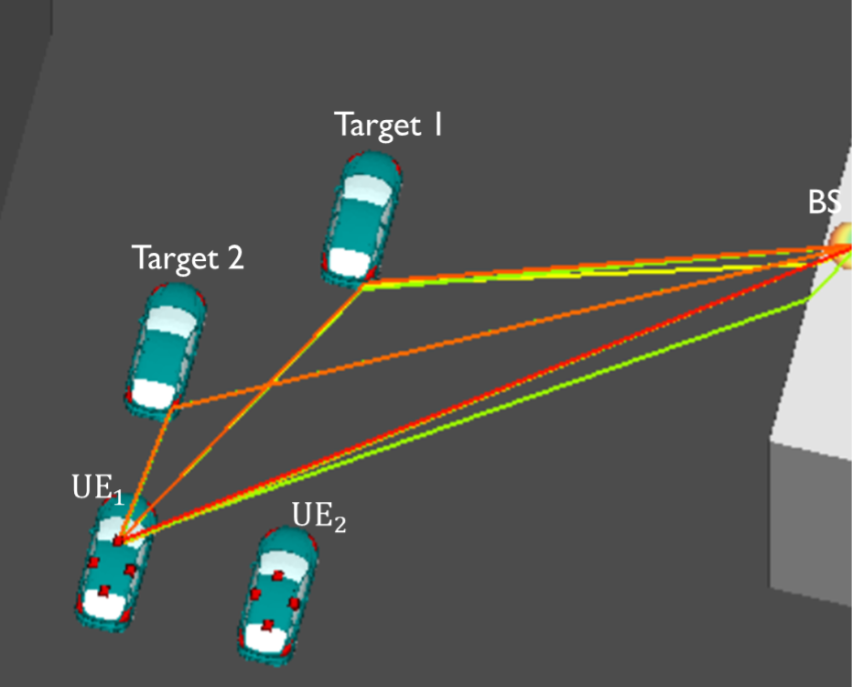}}
  \hfill
  \subfloat[\label{fig:pdp}]{%
    \includegraphics[width=0.24\textwidth]{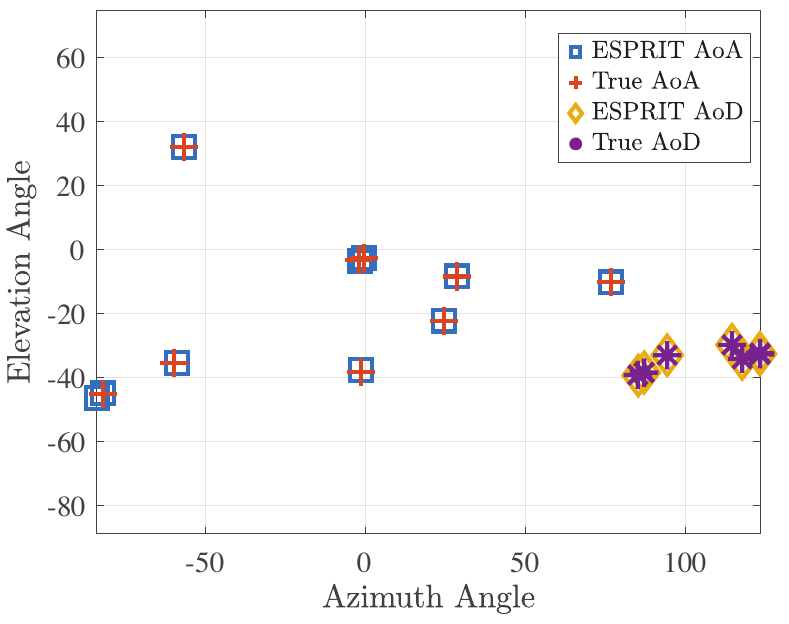}}
  \hfill
  \subfloat[\label{fig:ang}]{%
    \includegraphics[width=0.24\textwidth]{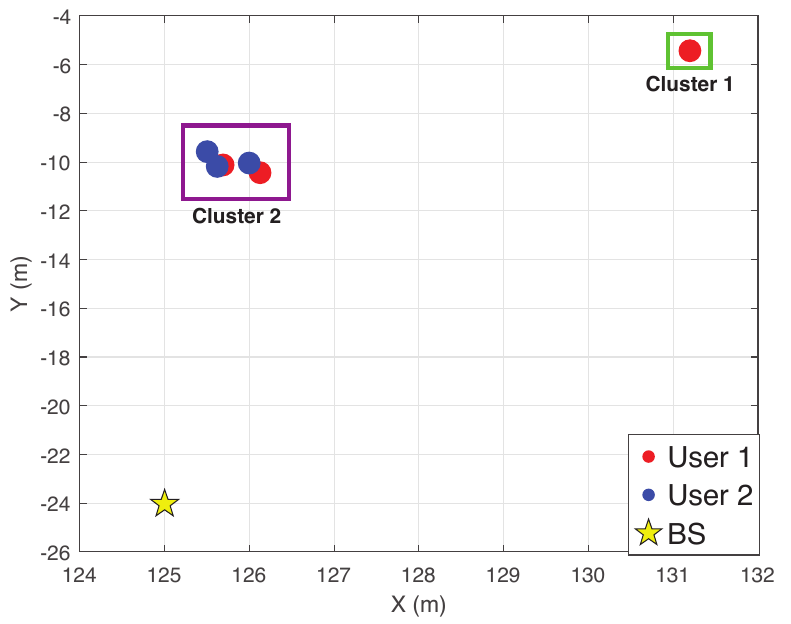}}
  \hfill
  \subfloat[\label{fig:tloc}]{%
    \includegraphics[width=0.24\textwidth]{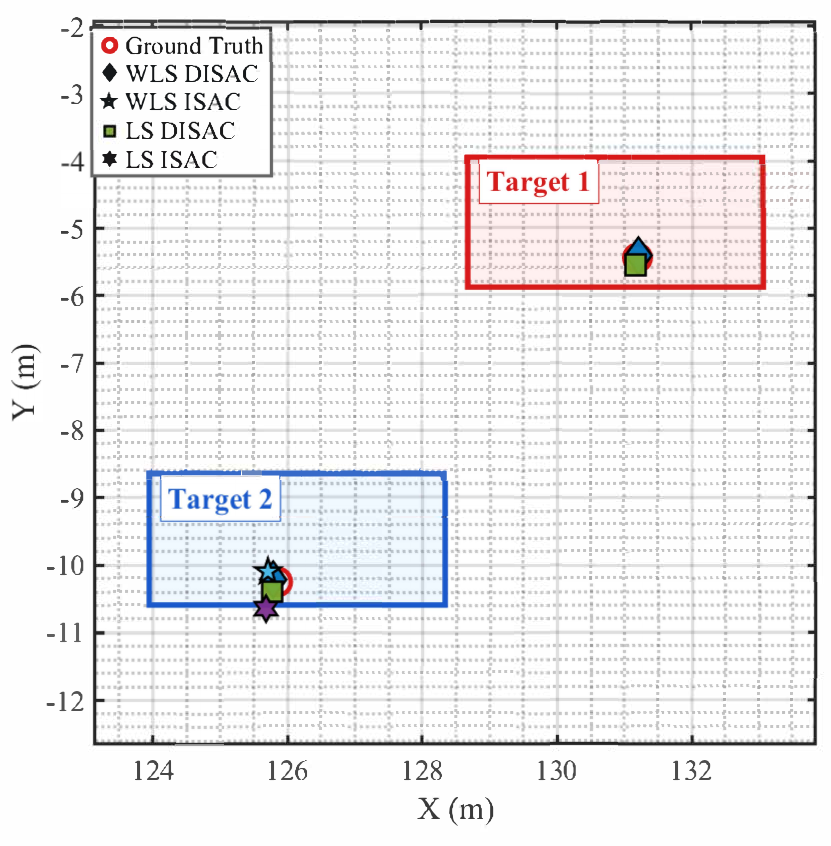}}
     \hfill
 % \subfloat[UE positioning/Target localization empirical CDF\label{fig:tloc}]{%
    %\includegraphics[width=0.24\textwidth,trim=12 12 12 12,clip]{Figs/CDF_Localization_Positioning_eps.eps}}
  \caption{Performance evaluation in a sample scenario: (a) UEs and targets  in the ray tracing environment simulated at 15 GHz. (b) Angular estimates of the multipath components vs. true values;  (d) Output of the clustering and association algorithm, which successfully clusters and labels all paths. (d) Target localization results vs. true values.}
  \label{fig:results}
  % \vspace{-3mm}
\end{figure}

\vspace*{-2mm}
We make use of the ray-tracing software Wireless-InSite to simulate a vehicular DISAC setting.  The BS is mounted on a building at $14$ m height. Both UEs and targets are hatchbacks. The carrier  frequency is set to 15 GHz, with 100MHz bandwidth. The transmit power is $P_t = 40$ dBm, and the noise variance is $\sigma^2=-93.85$ dBm. The BS is equipped with a $32\times 32$ UPA with $32$ RF chains, and the UEs are equipped with $8\times 8$ UPAs with 8 RF chains. 
We consider first the sample scenario with two UEs and two targets shown in Fig.~\ref{fig:results}(a). The channel estimation performance is illustrated in Fig.~\ref{fig:results}(b) by showing the estimated and true values of the AoAs and AoDs. We also show in Fig.~\ref{fig:results}(c) the output of the clustering and association algorithm, which identifies and labels the clusters without error. 
Finally, for this sample scenario we also show the target localization performance under various algorithms and scenarios. We consider two cases and two localization algorithms: DISAC with two collaborating UEs, single UE ISAC, and LS and WLS algorithms for localization. Fig.~\ref{fig:results}(d) shows how DISAC offers significant improvement over single UE ISAC, achieving sub-decimeter level localization accuracy when exploiting WLS. We observe that when the UEs do not collaborate, one target is not detected due to a lack of reflections, pointing out a clear limitation of single UE ISAC. The precise values of the target localization errors in Fig ~\ref{fig:results}(d)  are gathered in Table \ref{tab:per_target_errors}.

\begin{table}[h!]
\centering
\caption{Per-target localization errors (m)}
\label{tab:per_target_errors}
\setlength{\tabcolsep}{3pt}        % tighten column spacing
\renewcommand{\arraystretch}{0.95} % tighten row spacing
\footnotesize                      % smaller text
\begin{tabular*}{\linewidth}{@{\extracolsep{\fill}} r c c c c}
\hline
Target & WLS DISAC & WLS ISAC & LS DISAC & LS ISAC \\
\hline
1 & 0.075784 & N/A & 0.19728 & N/A \\
2 & 0.049561 & 0.24366  & 0.11852 & 0.53724 \\
\hline
\end{tabular*}
\end{table}

\begin{figure}[h!]
  \centering
    \includegraphics[width=0.35\textwidth]{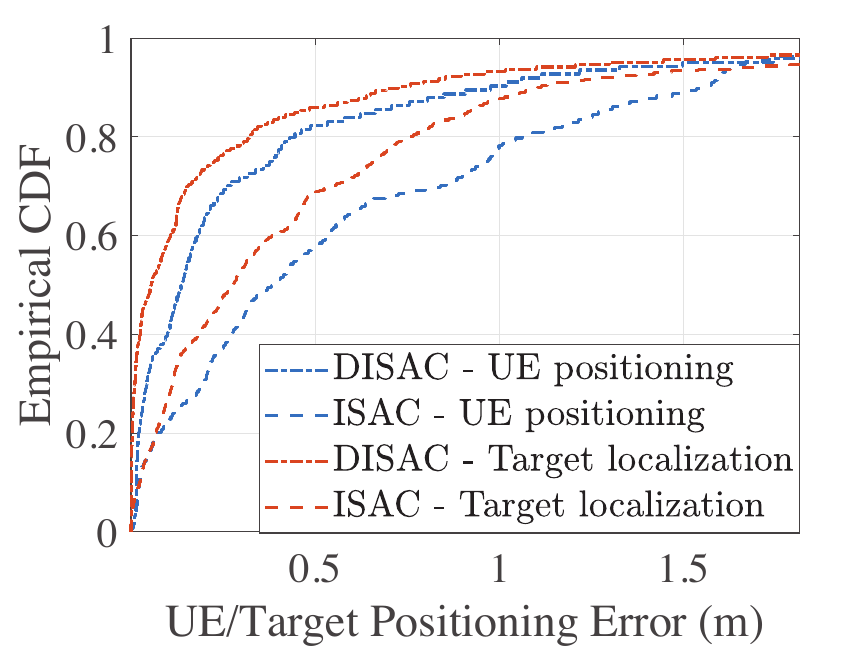}
    \caption{Cumulative distribution function for the UE and target localization errors considering a Monte Carlo simulation with 200 snapshots created by ray tracing.}
    \label{fig:cdf}    
    \end{figure}

To further validate the robustness of the system and the superiority of DISAC over single UE ISAC, we consider Monte-Carlo simulations over a dataset of 200 snapshots, with random locations of UEs and targets. We show the UE and target localization cumulative distribution function with DISAC and ISAC systems in Fig.~\ref{fig:cdf}. As we can see, DISAC offers substantial improvement over ISAC with a localization error below 32 cm for targets and below 43 cm for UE positioning for 80\% of the cases.  

\vspace*{-4mm}
\section{Conclusion}
\vspace*{-2mm}
This work proposed a single-snapshot DISAC framework for joint UE positioning and target localization in the upper mid-band. By leveraging distributed UEs, the system was shown to overcome the limitations of single-UE ISAC, even when operating with extended target models and realistic multipath channels. A novel processing chain incorporating multipath estimation, clutter removal, clustering, and WLS fusion enabled reliable joint estimation of UE positions, target locations, and timing offsets. Simulations based on ray-tracing channels demonstrated that DISAC achieves consistent decimeter-level accuracy for both UEs and targets, even under imperfect synchronization. These results highlight the potential of DISAC to unlock highly accurate multi-perspective sensing and positioning in FR3 systems.
%
%In this paper, we designed a single-snapshot DISAC pipeline for FR3 vehicular and target localization. The system model considers unsynchronized receivers, extended targets, and dense multipath. A beamspace ESPRIT algorithm extracts path angles, delays, and gains; clutter removal is accounted for by limiting the estimates to the FoI; and a linear fusion stage solves joint UE/target positioning and TO estimation via LS/WLS. Numerical experiments based on ray tracing channels show that cooperation between the UEs outperforms a single-UE ISAC baseline. In addition, decimeter-level localization accuracy for UEs and targets is guaranteed under imperfect synchronization.

\bibliographystyle{IEEEbib}
\bibliography{refs}

\end{document}